\documentclass[11pt]{article}
\usepackage{amsmath,amsthm}

\newcommand{\R}{{\mathbf R}} 
\newcommand{\N}{{\mathbf N}}
\newcommand{\K}{{\mathbf K}} 
\newcommand{\Z}{{\mathbf Z}}
\def\C{{\mathbf C}}
\newcommand{\Prm}{{\mathbf P}}

\renewcommand{\epsilon}{\varepsilon } 
 
\renewcommand{\rho}{\varrho } 
 
\renewcommand{\phi}{\varphi }

\renewcommand{\b}{\beta }

\def\rs{\right>}
\def\lg{\left|}

\newtheorem{theorem}{Theorem}
\newtheorem{lemma}{Lemma}
\newtheorem{corollary}{Corollary}
\newtheorem{proposition}{Proposition}

\begin {document}
\title{On a Problem in Quantum Summation}

\author {Stefan Heinrich\\
Fachbereich Informatik\\
Universit\"at Kaiserslautern\\
D-67653 Kaiserslautern, Germany\\
e-mail: heinrich@informatik.uni-kl.de
\\\\
and\\\\
Erich  Novak\\
Mathematisches Institut\\   
Universit\"at Jena\\
D-07740 Jena, Germany\\
e-mail: novak@mathematik.uni-jena.de}
\date{}
\maketitle

\date{}
\maketitle
\begin{abstract}
We consider the computation of the mean of sequences in the quantum model of
computation. We determine the query complexity in the case of
sequences which satisfy a $p$-summability condition for $1\le p<2$.
This settles a problem left open in Heinrich (2001).
\end{abstract}
\section{Introduction}
Computation of the mean of sequences and, equivalently, summation of sequences,
is an important numerical task, in particular for huge number of summands
occurring in many numerical applications such as, e.g., high dimensional
integration. The larger the number of summands (the larger the dimension),
the less these problems are tractable on a classical computer. It is therefore
an interesting and challenging task to understand to which extent
a quantum computer could bring speed-ups. First results for the 
summation of bounded sequences are due to Grover (1998), Nayak and Wu (1999),
Brassard, H{\o}yer, Mosca, and Tapp (2000). The case of sequences satisfying 
a $p$-summability condition, which arises in various problems like integration
of functions from $L_p$ and Sobolev classes, was studied in Heinrich (2001).
Up to logarithmic factors for $p=2$, in the case $2\le p <\infty$ 
the query complexity of the summation problem was determined. For the
case $1\le p<2$, matching upper and lower bounds were obtained only under
an additional restriction. The bounds for the remaining case did not match.
In this paper we settle this problem and determine the query complexity
in the full range of parameters.

Applications of our results to the quantum complexity of integration of functions from
Sobolev classes are given in Heinrich (2001a). The use of quantum summation 
for integration was first pointed out by Abrams and Williams (1999). 
The quantum complexity of 
integration was studied in Novak (2001), later in Heinrich (2001) and 
Heinrich and Novak (2001). Path integration is discussed in 
Traub and Wo\'zniakowski (2001). 
Furthermore, we refer to the surveys 
Ekert, Hayden, and Inamori (2000), Shor (2000), and to the monographs Pittenger (1999), 
Gruska (1999) and Nielsen and Chuang (2000) for general 
reading on quantum computation.

Our analysis is based on the framework introduced in Heinrich (2001)
of quantum algorithms for the approximate solution of problems of analysis.
This approach is an extension of the framework of information-based complexity
theory (see Traub, Wasilkowski, and Wo\'zniakowski, 1988, Novak, 1988, and, more 
formally, Novak, 1995)                           
to quantum computation. It also extends the binary black box model of 
quantum computation (see, e.g., Beals, Buhrman, 
Cleve, and Mosca, 1998) to situations where mappings from spaces of functions to 
the scalar field (such as the mean or the integral) have to be computed. 
Let us recall the main
notions here. For more details and background discussion we refer to
Heinrich (2001). 

Let $D$, $K$ be nonempty sets, let $\mathcal{F}(D,K)$
denote the set of all functions from $D$ to $K$, and let 
$F \subseteq \mathcal{F}(D, K)$ be a nonempty subset. 
Let $\K$, the scalar field, be either 
$\R$ or $\C$, the field of real or complex numbers, let  
$G$ be a normed space over $\K$, and 
let $S:F\to G$ be a mapping. We seek to 
approximate $S(f)$ for $f\in F$ by means of quantum computations. 
Let $H_1$ be the 
two-dimensional complex Hilbert space $\C^2$, with its unit vector
basis $\{e_0,e_1\}$, let
$$
 H_m=H_1\otimes\dots\otimes H_1 
$$
be the tensor product of $m$ copies of $H_1$, endowed with the tensor
Hilbert space structure. 
The following notation is convenient:
$$\Z[0,N) := \{0,\dots,N-1\}$$
for $N\in\N$ (as usual, $\N= \{1,2,\dots \}$ and $\N_0=\N\cup\{0\})$.
Let $\mathcal{C}_m = \{\lg i\rs:\, i\in\Z[0,2^m)\}$ be the canonical basis of
$H_m$, where  $\lg i \rs$ stands for 
$e_{j_0}\otimes\dots\otimes e_{j_{m-1}}$, $i=\sum_{k=0}^{m-1}j_k2^{m-1-k}$ the binary 
expansion of $i$. Denote the set of unitary operators on $H_m$ by $\mathcal{U}(H_m)$. 

A quantum query  on $F$ is given by a tuple
\begin{equation}
\label{J1}
Q=(m,m',m'',Z,\tau,\beta),
\end{equation}
where $m,m',m''\in \N, m'+m''\le m, Z\subseteq \Z[0,2^{m'})$ is a nonempty 
subset, and
$$\tau:Z\to D$$
$$\beta:K\to\Z[0,2^{m''})$$
are arbitrary mappings. Denote $m(Q):=m$, the number of qubits of $Q$. 

Given such a query $Q$, we define for each $f\in F$ the unitary operator 
$Q_f$ by setting for  
$\lg i\rs\lg x\rs\lg y\rs\in \mathcal{C}_m
=\mathcal{C}_{m'}\otimes\mathcal{C}_{m''}\otimes\mathcal{C}_{m-m'-m''}$:
\begin{equation}
\label{AB1} 
Q_f\lg i\rs\lg x\rs\lg y\rs=
\left\{\begin{array}{ll}
\lg i\rs\lg x\oplus\beta(f(\tau(i)))\rs\lg y\rs &\quad \mbox {if} \quad i\in Z\\
\lg i\rs\lg x\rs\lg y\rs & \quad\mbox{otherwise,} 
 \end{array}
\right. 
\end{equation}
where $\oplus$ means addition modulo $2^{m''}$. 

A quantum algorithm  on $F$  with no measurement is a tuple
\begin{equation*}
A=(Q,(U_j)_{j=0}^n),
\end{equation*}
where $Q$ is a quantum query on $F$, $n\in\N_0$  and
$U_{j}\in \mathcal{U}(H_m)\,(j=0,\dots,n)$, with $m=m(Q)$.
Given $f\in F$,
we let $A_f\in \mathcal{U}(H_m)$ be defined as
\begin{equation}
\label{B1a}
A_f = U_n Q_f U_{n-1}\dots U_1 Q_f U_0.
\end{equation}
We denote by $n_q(A):=n$ the number of queries and by $m(A)=m=m(Q)$ the 
number of qubits of $A$. Let $(A_f(x,y))_{x,y\in \Z[0,2^m)}$ 
be the matrix of the 
transformation $A_f$ in the canonical basis $\mathcal{C}_{m}$.

A quantum algorithm on $F$ with output 
in $G$ (or shortly, from $F$ to $G$) with $k$ measurements  is a tuple
$$
A=((A_\ell)_{\ell=0}^{k-1},(b_\ell)_{\ell=0}^{k-1},\varphi),
$$ 
where $k\in\N,$ and $A_\ell\,(\ell=0,\dots,k-1)$ are quantum algorithms
on $F$ with no measurements, 
$$
b_0\in\Z[0,2^{m_0}), 
$$
for $1\le \ell \le k-1,\,b_\ell$ is a function
$$
b_\ell:\prod_{i=0}^{\ell-1}\Z[0,2^{m_i}) \to \Z[0,2^{m_\ell}),
$$
where we denoted  $m_\ell:=m(A_\ell)$, and $\varphi$ is a function with values in $G$
$$
\varphi:\prod_{\ell=0}^{k-1}\Z[0,2^{m_\ell}) \to G.
$$
The output of $A$ at input $f\in F$ will be a probability measure $A(f)$ on $G$, 
defined as follows: First put
\begin{eqnarray}   
p_{A,f}(x_0,\dots, x_{k-1})&=&
|A_{0,f}(x_0,b_0)|^2 |A_{1,f}(x_1,b_1(x_0))|^2\dots\nonumber\\
&&\dots |A_{k-1,f}(x_{k-1},b_{k-1}(x_0,\dots,x_{k-2}))|^2.\label{M1}
\end{eqnarray}
Then define $A(f)$ by setting for any subset $C\subseteq G$
\begin{equation}
\label{M3}
A(f)(C)=\sum_{\phi(x_0,\dots,x_{k-1})\in C}p_{A,f}(x_0,\dots, x_{k-1}).
\end{equation}
By
$n_q(A):=\sum_{\ell=0}^{k-1} n_q(A_\ell)$
we denote the number of queries used by $A$.

Informally, such an algorithm $A$
starts with a fixed basis state $b_0$ and, at input $f$, applies in an alternating 
way unitary 
transformations $U_{0j}$ (not depending on $f$) and the operator $Q_f$
of a certain query. After a fixed
number of steps the resulting state is measured, which gives a (random) basis 
state, say $\xi_0$. This state is memorized and then transformed (e.g., by a classical
computation, which is symbolized by $b_1$) into a new basis state $b_1(\xi_0)$. 
This is the starting state to which the
next sequence of quantum operations is applied 
(with possibly another query and number of qubits). The resulting state is again measured, 
which gives the (random) basis state $\xi_1$. This state is memorized,  
$b_2(\xi_0,\xi_1)$ is computed (classically), and so on. After $k$ such cycles, we obtain
$\xi_0,\dots,\xi_{k-1}$. Then finally an element of $G$ 
is computed (e.g., again on a classical computer) from the results of all measurements:
$\varphi(\xi_0,\dots,\xi_{k-1})$. The probability measure $A(f)$ is its distribution.
For details, see Heinrich (2001).  

The error 
of $A$ is defined as follows: Let $0\le\theta< 1$, $f\in F$, and let 
$\zeta$ be any random variable with distribution $A(f)$. Then put
\begin{equation*}
e(S,A,f,\theta)=\inf\left\{\varepsilon\,\,|\,\,\Prm\{\|S(f)-\zeta\|>\varepsilon\}\le\theta
\right\}.
\end{equation*}
Associated with this we introduce
$$
e(S,A,F,\theta)=\sup_{f\in F} e(S,A,f,\theta), 
$$
$$
e(S,A,f)=e(S,A,f,1/4),
$$
and
$$
e(S,A,F)=e(S,A,F,1/4).
$$
The $n$-th minimal query error is defined for $n\in\N_0$ as
$$
e_n^q(S,F)=\inf\{e(S,A,F)\,\,|\,\,A\,\,
\mbox{is any quantum algorithm with}\,\, n_q(A)\le n\}.
$$  
This is the minimal error which can be reached using at most $n$ queries. 
The query complexity is defined for  
$\varepsilon > 0$ by 
\begin{eqnarray*}
\lefteqn{\mbox{comp}_\varepsilon^q(S,F)=}\\
&&\min\{n_q(A)\,\,|\,\, A\,\,\mbox{is any quantum 
algorithm with}\,\, e(S,A,F) \le \varepsilon\}.
\end{eqnarray*} 
The quantities $e_n^q(S,F)$ and $\mbox{comp}_\varepsilon^q(S,F)$ are inverse to each 
other in the following sense: 
For all $n\in \N_0$ and $\varepsilon > 0$,
$e_n^q(S,F)\le \varepsilon$ if and only if
$\mbox{comp}_{\varepsilon_1}^q(S,F)\le n$ for all $\varepsilon_1 > \varepsilon$.
Thus, determining the query complexity is equivalent to determining the
$n$-th minimal error. Henceforth, we will deal only with $e_n^q(S,F)$.

\section{The Main Result}
    Let $N\in\N$ and set $D=\Z[0,N)$, $K=\R$, $G=\R$. For  $1\le p\le\infty$ 
let $L_p^N$ denote the space of all functions
$f:D\to \R$, equipped with the norm 
$$
\|f\|_{L_p^N}=\left(\frac{1}{N}\sum_{i=0}^{N-1}|f(i)|^p \right)^{1/p}
$$
if $p<\infty$
and
$$
\|f\|_{L_\infty^N}=\max_{0\le i\le N-1} |f(i)|.
$$
Define $S_N:L_p^N\to\R$ by 
$$
S_N f=\frac{1}{N}\sum_{i=0}^{N-1}f(i) 
$$  and let
$$
F=\mathcal{B}_p^N:=\{f\in L_p^N \,|\, \|f\|_{L_p^N}\le 1\}.
$$

Let us summarize the known results about the order of 
$e_n^q(S_N,\mathcal{B}_p^N)$ (and thus the query complexity of computing 
the mean of $p$-summable sequences) in Theorem \ref{theo:3}. 
The case $p=\infty$ is due to
Grover (1998), Brassard, H{\o}yer, Mosca, and Tapp (2000) (upper bounds)
and Nayak and Wu (1999) (lower bounds).
The results in the case $1\le p<\infty$ are due to Heinrich (2001). 
Note that throughout the paper we often use the same symbols for
possibly different constants. Also,   $\log$  always means $\log_2$.

\begin{theorem}
\label{theo:3}
Let $1\le p\le\infty$. There are constants $c_0,c_1,c_2,c_3>0$ 
such that for all $n,N\in\N$ with $2<n\le c_1 N$,
$$
c_2 n^{-1}\le e_n^q(S_N,\mathcal{B}_p^N)\le c_3 n^{-1}\quad\mbox{if}\quad 2<p\le\infty,
$$
$$
c_2 n^{-1}\le e_n^q(S_N,\mathcal{B}_2^N)\le c_3 n^{-1}\log^{3/2}n\log\log n,
$$
and 
$$
c_2 n^{-2(1-1/p)}\le e_n^q(S_N,\mathcal{B}_p^N)\le c_3 n^{-2(1-1/p)}
\quad\mbox{if}\quad 1\le p<2,\,\,n\le c_0\sqrt{N}.
$$
\end{theorem}
The case $1\le p<2, \,\, n\ge c_0\sqrt{N}$ was left open.
We will settle it here by proving
\begin{theorem}
\label{theo:4}
Let $1\le p<2$. There are constants $c_0,c_1,c_2,c_3>0$ 
such that for all $n,N\in\N$ with $c_0\sqrt{N}\le n\le c_1 N$,   
$$
c_2 n^{-2/p}N^{2/p-1}\le e_n^q(S_N,\mathcal{B}_p^N)\le c_3n^{-2/p}N^{2/p-1}
\max(\log(n/\sqrt{N}),1)^{2/p-1}.
$$ 
\end{theorem}
It is interesting to mention the consequences for the case $p=1$
separately:
\begin{corollary}
\label{cor:2}
There are constants $c_1,c_2,c_3>0$ such that 
$$
c_2\le e_n^q(S_N,\mathcal{B}_1^N) \le 1 
$$
if $ 0\le n<\sqrt{N}$, and 
$$
c_2n^{-2}N\le e_n^q(S_N,\mathcal{B}_1^N)\le c_3n^{-2}N\max(\log(n/\sqrt{N}),1)
$$
if $\sqrt{N}\le n\le c_1 N$.
\end{corollary}
Hence the decay 
essentially starts only beyond $\sqrt{N}$. Note
that the corresponding quantities for the classical deterministic  and randomized
setting remain $\Omega(1)$ also in the range $\sqrt{N}\le n\le c_1 N$, see Heinrich 
and Novak (2001).

Combining this with the respective result in Theorem \ref{theo:3},
we can cover the full range $n\le c_1 N$. This result is a direct consequence of Theorems
\ref{theo:3} and \ref{theo:4} and the monotonicity of $e_n^q(S_N,\mathcal{B}_p^N)$
in $n$.

\begin{corollary}
\label{cor:3}
Let $1\le p<2$. There are constants $c_1,c_2,c_3>0$ 
such that for all $n,N\in\N$ with $n\le c_1 N$,   
\begin{eqnarray*}
\lefteqn{
c_2 \min(n^{-2(1-1/p)},n^{-2/p}N^{2/p-1})\le e_n^q(S_N,\mathcal{B}_p^N)}\\
&&
\le c_3\min(n^{-2(1-1/p)},n^{-2/p}N^{2/p-1})\max(\log(n/\sqrt{N}),1)^{2/p-1}.
\end{eqnarray*}
\end{corollary} 

The following two sections contain the proof of Theorem \ref{theo:4}. 

\section{Upper Bounds}
For any $M\in\N$ we define
$$
S_{N,M}f=\frac{1}{N}\sum_{i\in\Z[0,N),\,|f(i)|<M}f(i)
$$
and
$$
S'_{N,M}f=S_Nf-S_{N,M}f=\frac{1}{N}\sum_{i\in\Z[0,N),\,|f(i)|\ge M}f(i).
$$

\begin{proposition}\label{pro:1}
Let $1\le p <\infty$. Then there is a constant $c>0$ such that for all $n,M,N\in\N$ 
with
$$
n\ge c\,M^{-p/2}N\max(\log(M^{-p}N),1)
$$
we have
$$
e_n^q(S'_{N,M},\mathcal{B}_p^N)=0.
$$
\end{proposition}
\begin{proof}
It is easily verified that 
$$
e_N^q(S'_{N,M},\mathcal{B}_p^N)=0
$$
(we use the queries just classically to obtain the values of the $f(i)$
up to any required precision and compute the sum classically). It follows that, modifying
$c$, if necessary, it suffices to prove the result for 
\begin{equation}
\label{A1}
M\ge M_0.
\end{equation}
We will specify $M_0$ later on. Furthermore, we may also asssume that
\begin{equation}
\label{A2}
M^p\le N,
\end{equation}
because otherwise $S'_{N,M}f=0$ for all $f\in \mathcal{B}_p^N$, so $e_0^q(S'_{N,M})=0$.
Let
\begin{equation}
\label{A3}
m'=\lceil\log N\rceil.
\end{equation}
First we define a quantum algorithm $A_0$ from $\mathcal{B}_p^N$ to 
$\Z[0,2^{m'})\times\R$. To specify its quantum query, fix
any $m''>m'+1$ and define the mapping $\b:\R\to\Z[0,2^{m''})$ by setting for $z\in\R$
\[
\b(z)=
\left\{\begin{array}{lll}
   2^{m''-1}       & \mbox{if} \quad |z|<M   \\
   \lfloor 2^{m''-m'-1}(z+2^{m'})\rfloor       & \mbox{if} \quad  M\le|z|<2^{m'}\\
   2^{m''}-1& \mbox{if} \quad z\ge 2^{m'} \\
   0 & \mbox{if} \quad z\le -2^{m'}.
   \end{array}
   \right.
\]
It follows that for $M\le |z|\le2^{m'}$,
\begin{equation}
\label{A4}
-2^{m'}+2^{-m''+m'+1}\b(z)\le z \le -2^{m'}+2^{-m''+m'+1}(\b(z)+1),
\end{equation}
and 
\begin{equation}
\label{A5}
\b(z)=2^{m''-1} \quad\mbox{if and only if}\quad |z|<M.
\end{equation}
In connection with this definition let us mention that for $f\in\mathcal{B}_p^N$,
\begin{equation}
\label{A6}
|f(i)|\le N^{1/p}\le N \le 2^{m'} \quad (i=0,\dots,N-1).
\end{equation}
Put $Z=\Z[0,N)$, let $\tau:Z\to\Z[0,2^{m'})$ be the identical embedding, $m=m'+m''$, and
define the query by
$$
Q=(m,m',m'',Z,\tau,\b).
$$
Let $H_m=H_{m'}\otimes H_{m''}$, and let
$$
\lg i\rs\lg x\rs\quad (i\in Z[0,2^{m'}),\,x\in Z[0,2^{m''}))
$$
be the respective representation of basis states. Let $W_0\in\mathcal{U}(H_{m'})$ 
be the Walsh-Hadamard transform, and let $X_0\in\mathcal{U}(H_{m'})$ be defined by
$$
X_0\lg i\rs=\left\{\begin{array}{rll}
  -\lg i\rs &\mbox{if} \quad i=0 \\
  \lg i\rs & \mbox{otherwise.}    \\
    \end{array}
\right.
$$
Consider the following unitary transforms on $H_m$, defined by: 
\begin{eqnarray*}
W\lg i\rs\lg x\rs&=&(W_0\lg i\rs)\lg x\rs,\\
X\lg i\rs\lg x\rs&=&
(X_0\lg i\rs)\lg x\rs,\\
T\lg i\rs\lg x\rs &=&
\left\{\begin{array}{rll}
\lg i\rs\lg x\rs&  \mbox{if} \quad i\in Z\,\,\mbox{and}\,\,x\ne 2^{m''-1}\\
  -\lg i\rs\lg x\rs&\mbox{otherwise,}  \\
    \end{array}
\right.\\
J\lg i\rs\lg x\rs&=&\lg i\rs\lg \ominus x\rs.
\end{eqnarray*}
Here $\ominus x$ stands for $(2^{m''}-x)\,\mbox{mod}\,2^{m''}$.
Note that $W_0^{-1}=W_0$, and hence $W^{-1}=W$.
For $f\in \mathcal{B}_p^N$ put 
\begin{equation}
\label{A7}
Y_f=WXWQ_f JT Q_f.
\end{equation}
Denote
$$
D_f=\big\{i\,\big|\,i\in Z,\,|f(i)|\ge M\big\}.
$$ 
It follows from the definitions above and from (\ref{A5}) that
\[
Q_f JT Q_f\lg i\rs\lg 0\rs=
\left\{\begin{array}{rll}
   \lg i\rs\lg 0\rs & \mbox{if} \quad i\in D_f   \\
  -\lg i\rs\lg 0\rs & \mbox{otherwise.}    \\
    \end{array}
\right. 
\]
$A_0$ will be an algorithm with one measurement. We define its unitary transform as 
\begin{equation}
\label{A8}
Q_f Y_f^L W,
\end{equation}
where $L\in\N$ will be specified later. The starting state will be 
$\lg b_0\rs=\lg 0\rs\lg 0\rs$, and the 
mapping $\phi:\Z[0,2^{m'})\times\Z[0,2^{m''})\to \Z[0,2^{m'})\times\R$
will be given by
\begin{equation}
\label{A9}
\phi(i,x)=(i,-2^{m'}+2^{-m''+m'+1}x).
\end{equation}
This completes the definition of algorithm $A_0$. Clearly, $Y_f$ is the
Grover iterate for the set $D_f$,
and the whole algorithm is Grover's search algorithm (Grover, 1996), or amplitude 
amplification, in the terminology
of Brassard, H{\o}yer, Mosca, and Tapp (2000), with respect to the $H_{m'}$ component,
followed by one more query $Q_f$. Observe that by (\ref{A4}) and 
(\ref{A6}) each run of the 
algorithm $A_0$ produces a pair $(i,y)\in \Z[0,2^{m'})\times\R$ with
\begin{equation}
\label{A10}
y\le f(i)\le y+2^{-m''+m'+1}\quad\mbox{if}\quad i\in D_f
\end{equation}
and
\begin{equation}
\label{A11}
y=0\quad\mbox{if and only if}\quad i<N\,\,\mbox{and}\,\,i\not\in D_f.
\end{equation}
The final algorithm $A$ is defined as $\psi(A_0^{L^*})$, which means that we repeat
$A_0$ $L^*$ times and compose the outputs by the mapping 
$$
\psi:(\Z[0,2^{m'})\times\R)^{L^*}\to\R,
$$
see Heinrich (2001), Section 2, for a formal definition. The number $L^*\in\N$ will
be specified later. The mapping $\psi$ is defined as follows: 
Let
$$
(i_\ell,y_\ell)_{\ell=0}^{L^*-1}\in (\Z[0,2^{m'})\times\R)^{L^*}
$$
be the outputs of the $L^*$ runs of $A_0$. We exclude all pairs with
$i_\ell\not\in D_f$ (which amounts to checking if $i\ge N$ or $y=0$, by (\ref{A11})), 
as well as all repetitions of any
$i_\ell\in D_f$ (by a suitable sorting algorithm). For the remaining set we add 
the second components and divide by $N$ (if the remaining set is empty, we output 0).

Now we show that with a suitable choice of the parameters $m'',L,L^*$, the algorithm 
outputs $S'_{N,M}f$ with error at most $2^{-m''+m'+1}$ with probability
at least $3/4$. This follows from (\ref{A10}) if we prove that with probability
at least 3/4 the set of remaining indices equals $D_f$. If $D_f=\emptyset$, this
is trivial, so we assume $D_f\ne\emptyset$. First we analyze $A_0$.
Denote $\mu_f=|D_f|$, hence $\mu_f\ge 1$, and let $0<\theta_f\le \pi/2$ be defined by
\begin{equation}
\label{B1}
\sin^2\theta_f=2^{-m'}\mu_f.  
\end{equation}
Finally, let
$$
\lg\psi_{f,1}\rs=2^{-m'/2}\sum_{i\in D_f}\lg i\rs
$$
and
$$
\lg\psi_{f,0}\rs=2^{-m'/2}\sum_{i\in \Z[0,2^{m'})\setminus D_f}\lg i\rs.
$$
By the analysis of Brassard, H{\o}yer, Mosca, and Tapp (2000), relation (8),
\begin{eqnarray*}\lefteqn{   
Y_f^LW\lg 0\rs\lg 0\rs
=(2^{-m'}\mu_f)^{-1/2}\sin((2L+1)\theta_f)\lg\psi_{f,1}\rs\lg 0\rs+}\\
&&\hspace {2.5cm}(1-2^{-m'}\mu_f)^{-1/2}\cos((2L+1)\theta_f)\lg\psi_{f,0}\rs\lg 0\rs,
\end{eqnarray*}
(where the second term is replaced by $0$ if $\mu_f=2^{m'}$). It follows that for any
$i_0\in D_f$, the algorithm $A_0$ outputs $(i_0,\b(f(i_0)))$ with probability
\begin{equation}
\label{B4}
\rho_{i_0}=\mu_f^{-1}\sin^2((2L+1)\theta_f).
\end{equation}
In the sequel we use the elementary relation 
\begin{equation}
\label{A12}
2x/\pi\le\sin x\le x \quad(x\in [0,\pi/2]).
\end{equation}
Since $f\in\mathcal{B}_p^N$, we have 
$$
N^{-1}M^p|D_f|\le 1,
$$
hence
\begin{equation}
\label{C1}
\mu_f=|D_f|\le M^{-p}N
\end{equation}
and $$
2^{-m'}\mu_f\le M^{-p}N\,2^{-m'}\le M^{-p}.
$$
Therefore, by (\ref{A12}) and (\ref{B1})
$$
4\pi^{-2}\theta_f^2\le M^{-p}
$$
and hence
\begin{equation}
\label{A20}
\theta_f\le 2^{-1}\pi M^{-p/2}.
\end{equation}
Now we put 
\begin{equation}
\label{F1}
M_0=\lceil6^{2/p}\rceil
\end{equation}
and define $L$ by
\begin{equation}
\label{B2}
L=\lfloor 3^{-1}M^{p/2}\rfloor.
\end{equation}
Since we assumed $M\ge M_0$, we get from (\ref{F1}) and (\ref{B2}), 
\begin{equation}
\label{C2}
1\le\frac{1}{6}M^{p/2}\le L\le \frac{1}{3}M^{p/2}.
\end{equation}
It follows from (\ref{A20}) and (\ref{C2}) that 
\begin{equation}
\label{B5}
(2L+1)\theta_f\le 3L\theta_f\le\pi/2.
\end{equation}
On the other hand, by (\ref{C2}) and (\ref{B1}),
$$
(2L+1)\theta_f> 2L\theta_f\ge \frac{1}{3}M^{p/2}\sin\theta_f
= \frac{1}{3}M^{p/2}(2^{-m'}\mu_f)^{1/2}.
$$
 From (\ref{B4}), (\ref{A12}), (\ref{B5}) and the relation above,
\begin{eqnarray*}
\rho_{i_0}&\ge& \frac{4}{\pi^{2}}\mu_f^{-1}(2L+1)^2\theta_f^2\\
&\ge&\frac{4}{9\pi^2}M^p\,2^{-m'}\\
 &\ge&\frac{2}{9\pi^2}M^p N^{-1}=c_2M^pN^{-1},
\end{eqnarray*}
where in the last line we used (\ref{A3}) and set $c_2=2/(9\pi^2)$. 
It follows that after $L^*$ repetitions
of algorithm $A_0$ the probability of $(i_0,\b(f(i_0)))$ not being among the results
is 
$$
\le (1-c_2M^pN^{-1})^{L^*}\le e^{-c_2M^pN^{-1}L^*},
$$
where we used that $1+x\le e^x$ for $x\in \R$. The probability that at least
one $i_0\in D_f$ is not among the results is 
$$
\le \mu_f e^{-c_2M^pN^{-1}L^*}\le M^{-p}N e^{-c_2M^pN^{-1}L^*},
$$
where we used (\ref{C1}). Now we choose $L^*$ in such a way that this probability
is not greater than $1/4$. This requires (recall that $\log$ means $\log_2$)
$$
(c_2\log e)\,M^pN^{-1}L^*\ge\log (M^{-p}N)+2,
$$
which is satisfied if
$$
L^*=\left\lceil\frac{3}{c_2\log e}M^{-p}N\max(\log(M^{-p}N),1)\right\rceil.
$$
We put $c_3=3/(c_2\log e)$ and observe that the above combined with 
(\ref{A2}) implies
$$
L^*\le (c_3+1)M^{-p}N\max(\log(M^{-p}N),1).
$$
Together with (\ref{C2}), this implies that algorithm $A$ makes 
$$
(2L+1)L^*\le 3LL^*\le (c_3+1)M^{-p/2}N\max(\log(M^{-p}N),1)
$$
queries to compute $S'_{N,M}f$ up to error $2^{-m''+m'+1}$
 with probability at least 3/4. Since $m''$ was arbitrary, the result follows.
\end{proof}
We need to express $M$ in terms of $n$ and $N$:
\begin{corollary}\label{cor:1} Let $1\le p<\infty$.
There is a constant $c\ge 1$ such that for all $n,M,N\in\N$,
$$
e_n^q(S'_{N,M},\mathcal{B}_p^N)=0
$$
whenever
$$
M\ge c(N/n)^{2/p}\max(\log(n/\sqrt{N}),1)^{2/p}.
$$
\end{corollary}
\begin{proof}
Let $c_0$ be the constant from Proposition \ref{pro:1}.
We put
\begin{equation}
\label{G3}
c=\max((2c_0)^{2/p},1).
\end{equation}
Assume
$$
M\ge c(N/n)^{2/p}\max(\log(n/\sqrt{N}),1)^{2/p}.
$$
It follows that 
\begin{equation}
\label{G1}
M^{-p/2}N\le c^{-p/2}n/\max(\log(n/\sqrt{N}),\,1).
\end{equation}
Squaring and dividing by $N$ gives
$$
M^{-p}N\le c^{-p}n^2N^{-1}/\max(\log(n/\sqrt{N}),1)^2,
$$
and hence
\begin{eqnarray}
\lefteqn{
\max(\log(M^{-p}N),1)}\nonumber\\
&\le& 
\max\Big(\log(c^{-p})+2\log(n/\sqrt{N})
-2\log\big(\max(\log(n/\sqrt{N}),1)\big),1\Big)\nonumber\\
&\le&2\max(\log(n/\sqrt{N}),1). \label{G2}
\end{eqnarray}
(\ref{G1}), (\ref{G2}) and (\ref{G3}) give 
$$
c_0M^{-p/2}N\max(\log(M^{-p}N),1)\le 2c_0c^{-p/2}n\le n,
$$
which, by Proposition \ref{pro:1}, implies
$$
e_n^q(S'_{N,M},\mathcal{B}_p^N)=0.
$$
\end{proof}
\begin{proposition}\label{pro:2}
Let $1\le p<2$. There is a constant $c>0$ such that for all $k,n,N\in\N$,
$$
e_n^q(S_{N,2^k},\mathcal{B}_p^N)\le c(2^{(1-p/2)k}n^{-1}+2^kn^{-2}).
$$
\end{proposition}
\begin{proof}
This is a direct consequence of the method of  proof of Theorem 1 in Heinrich (2001). 
For the sake of completeness, we recall some key steps.
Since trivially $e_n^q(S_{N,2^k},\mathcal{B}_p^N)\le 1$ for all $n\in\N_0$ 
(just use the zero algorithm), it
suffices to prove the result under the assumption 
\begin{equation}
\label{L1}
n\ge 2^{(1-p/2)k}.
\end{equation}
Define $S_N^{\ell,\sigma}: L_p^N\to\R$ for $\ell=0,\dots, k,\,\sigma=0,1$ as
$$
S_N^{\ell,\sigma}f=(-1)^\sigma 2^{-\ell}N^{-1}
\sum_{2^{\ell-1}\le(-1)^\sigma f(i)<2^\ell}f(i)
$$
if $\ell\ge 1$ and
$$
S_N^{0,\sigma}f=(-1)^\sigma N^{-1}
\sum_{0\le (-1)^\sigma f(i)<1}f(i).
$$
 It is shown in Heinrich (2001) (based on the counting algorithm of 
Brassard, H{\o}yer, Mosca, and Tapp, 2000), that there is a constant $c>0$ such that 
for each choice of 
$\nu_\ell,\,n_\ell\in\N \,\,(\ell=0,\dots,k)$, there are algorithms 
$A_{\ell,\sigma}\,\,
(\ell=0,\dots,k$, $\sigma=0,1)$  with 
$n_q(A_{\ell,\sigma})\le \nu_\ell n_\ell$ and
$$
e(S_N^{\ell,\sigma},A_{\ell,\sigma},\mathcal{B}_p^N,2^{-\nu_\ell})
\le c(2^{-p\ell/2}n_\ell^{-1}+n_\ell^{-2})
$$
(use the relation following (27) in Heinrich, 2001, together with (21) 
and (22) of that paper).
Now choose
\begin{equation*}
n_\ell=\left\lceil 2^{-(1/2-p/4)(k-\ell)}n\right\rceil,
\end{equation*}
and
$$
\nu_\ell=\lceil2\log(k-\ell+1)\rceil+4.
$$
Due to (\ref{L1}),
\begin{equation}
\label{L2}
n_\ell< 2^{-(1/2-p/4)(k-\ell)+1}n.
\end{equation}
Let the algorithm $A$ be defined by
$$
A=\sum_{0\le\ell \le k \atop \sigma=0,1}(-1)^\sigma 2^\ell A_{\ell,\sigma}.
$$
(We refer again to Heinrich, 2001, Section 2, for a formal definition.)
Taking into account (\ref{L2}), it follows that
\begin{equation}
\label{H1}
n_q(A)\le 2\sum_{\ell=0}^k (\lceil2\log(k-\ell+1)\rceil+4)
\left\lceil 2^{-(1/2-p/4)(k-\ell)}n\right\rceil\le c_1 n.
\end{equation}
Moreover, since
$$
2\sum_{\ell=0}^k 2^{-\nu_\ell} 
\le \frac{1}{8}\sum_{\ell=0}^k (k-\ell+1)^{-2}<
\frac{1}{4},
$$
we get
\begin{eqnarray}
\lefteqn{
e(S_{N,2^k},A,\mathcal{B}_p^N)
}\nonumber\\
&\le & 
c\sum_{\ell =0}^k \left(2^{(1-p/2)\ell+(1/2-p/4)(k-\ell)}n^{-1} 
+2^{\ell+(1-p/2)(k-\ell)} n^{-2}\right)\nonumber\\
&\le &
c\sum_{\ell =0}^k \left(2^{(1/2-p/4)(k+\ell)}n^{-1} 
+2^{k-p(k-\ell)/2} n^{-2}\right)\nonumber\\
&\le &
c_2\left(2^{(1-p/2)k}n^{-1}+2^k n^{-2}\right)\nonumber
\end{eqnarray}
which together with (\ref{H1}) and a suitable scaling of $n$ implies the desired 
result.
\end{proof} 
\begin{theorem}
\label{theo:1}
Let $1\le p < 2$. There are constants $c_0,c>0$ such that for all $n,N\in\N$ with 
$n\ge c_0\sqrt{N}$
$$
e_n^q(S_N,\mathcal{B}_p^N)\le cn^{-2/p}N^{2/p-1}
\max(\log(n/\sqrt{N}),1)^{2/p-1}.
$$
\end{theorem}
\begin{proof}
First note that 
\begin{equation}
\label{N1}
e_N^q(S_N,\mathcal{B}_p^N)=0.
\end{equation}
Next observe that it follows readily from Lemma 3 in Heinrich (2001) 
(reducing the error probability
by repeating the algorithm and computing the median)
that there is a constant $c_0\in\N$ such that for all $n,k,N\in\N$,
\begin{equation}
\label{H3} 
e_{c_0n}^q(S_N,\mathcal{B}_p^N)
\le e_n^q(S_{N,2^k},\mathcal{B}_p^N) + e_n^q(S'_{N,2^k},\mathcal{B}_p^N).
\end{equation}
Now let $n$ satisfy
\begin{equation}
\label{N2}
\sqrt{N}\le n<N
\end{equation}
and choose $k\in\N$ in such a way that
$$
2^{k-1}< c_1(N/n)^{2/p}\max(\log(n/\sqrt{N}),1)^{2/p}\le 2^k,
$$
where $c_1\ge 1$ is the constant from Corollary \ref{cor:1}. Consequently, we have
\begin{equation}
\label{H4}
e_n^q(S'_{N,2^k},\mathcal{B}_p^N)=0.
\end{equation}
Moreover, with $c_2$ being the constant from Proposition \ref{pro:2},
\begin{eqnarray}
\label{H2}
\lefteqn{
e_n^q(S_{N,2^k},\mathcal{B}_p^N)}\nonumber\\
&\le& 
c_2\big(2^{(1-p/2)k}n^{-1}+2^kn^{-2}\big)\nonumber\\
&\le&
c_3\left((N/n)^{\frac{2}{p}(1-p/2)}n^{-1}
\max\left(\log\frac{n}{\sqrt{N}},1\right)^{\frac{2}{p}(1-p/2)}\right.\nonumber\\
&&\left.\quad+(N/n)^{2/p}n^{-2}
\max\left(\log\frac{n}{\sqrt{N}},1\right)^{2/p}\right)\nonumber\\
&=&
c_3\left(N^{2/p-1}n^{-2/p}
\max\left(\log\frac{n}{\sqrt{N}},1\right)^{2/p-1}\right.\nonumber\\
&&\left.\quad+N^{2/p}n^{-2/p-2}\max\left(\log\frac{n}{\sqrt{N}},1\right)^{2/p}\right).
\end{eqnarray}
Using (again) $x\ge \ln(1+x)$ for $x>-1$, we have
$$
\frac{n^2}{N}\ge\ln\left(\frac{n^2}{N}+1\right)\ge 2\ln\frac{n}{\sqrt{N}}=
\frac{2}{\log e}\log\frac{n}{\sqrt{N}}>\log\frac{n}{\sqrt{N}}.
$$
Consequently, recalling our assumption $n\ge\sqrt{N}$, we get
$$
\frac{n^2}{N}\ge \max\left(\log\frac{n}{\sqrt{N}},1\right),
$$
and therefore
$$
N^{2/p-1}n^{-2/p}\max\left(\log\frac{n}{\sqrt{N}},1\right)^{2/p-1}
\ge N^{2/p}n^{-2/p-2}\max\left(\log\frac{n}{\sqrt{N}},1\right)^{2/p}.
$$
 From (\ref{H3}), (\ref{H4}), (\ref{H2}), and the relation above we get
\begin{eqnarray}
e_{c_0 n}^q(S_N,\mathcal{B}_p^N)&\le &e_n^q(S_{N,2^k},\mathcal{B}_p^N)\nonumber\\
&\le& 2c_3 N^{2/p-1}n^{-2/p}\max\left(\log\frac{n}{\sqrt{N}},1\right)^{2/p-1}\label{H5}
\end{eqnarray}
for all $n$ with $\sqrt{N}\le n<N$. With a suitable scaling of $n$, the result 
follows from (\ref{H5}) and (\ref{N1}).
\end{proof}
\section{Lower Bounds}
We need some general results from
Section 4 of Heinrich (2001).
Let $D$ and $K$ be nonempty sets, let
$L\in \N$, and let to each $u=(u_0,\dots,u_{L-1})\in\{0,1\}^L$ an 
$f_u\in \mathcal{F}(D,K)$ be assigned such that the following 
is satisfied:
\\ \\
{\bf Condition (I):} For each $t\in D$ there is an $\ell$, $0\le \ell\le L-1$, 
such that $f_u(t)$ depends only on $u_\ell$, in other words, for $u,u'\in\{0,1\}^L$, 
$u_\ell=u'_\ell$ implies $f_u(t)=f_{u'}(t)$.
\\ \\ 
Define the function $\rho (L,\ell,\ell')$
for $L\in\N$, $0\le\ell\ne\ell'\le L$ by
\begin{equation}
\label{AC1}
\rho (L,\ell,\ell')=\sqrt{\frac{L}{|\ell-\ell'|}}+
\frac{\min_{j=\ell,\ell'}\sqrt{j(L-j)}}{|\ell-\ell'|}.
\end{equation}
The following was proved in Heinrich (2001), using the polynomial method of 
Beals, Buhrman, Cleve, and Mosca (1998) and based on a result of
Nayak and Wu (1999):
\begin{lemma}
\label{lem:5} There is a constant $c_0>0$ such that the following holds:
Let $D,K$ be nonempty sets, let 
$F\subseteq\mathcal{F}(D,K)$ be a set of functions, $G$ a normed space,
$S:F\to G$ a function, and $L\in\N$. Suppose 
$(f_u)_{u\in\{0,1\}^L}\subseteq\mathcal{F}(D,K)$ is a system of functions satisfying
condition (I). Let finally $0\le\ell\ne\ell'\le L$ and assume that 
\begin{equation}
\label{AC2}
f_u\in F \quad {\rm whenever} \quad
|u|\in\{\ell,\ell'\}.
\end{equation}
 Then
\begin{equation}
\label{C3}
e_n^q(S,F)\ge \frac{1}{2}\min\big\{ \|S(f_u)-S(f_{u'})\|\,\big |\, |u|=\ell,\, 
|u'|=\ell'\big\}
\end{equation}
for all $n$ with
\begin{equation}
\label{C4}
n\le c_0\rho (L,\ell,\ell').
\end{equation}
\end{lemma}
The next result contains lower 
bounds matching the upper ones from Theorem \ref{theo:1} 
up to a logarithmic factor.
\begin{theorem}
\label{theo:2}
Let $1 \le p <2$. 
Then there are constants $c_0,c_1,c_2 > 0$ such that for all $n, N \in \N$  with
 $c_0\sqrt{N}\le n\le c_1 N$,   
$$
e_n^q(S_N,\mathcal{B}_p^N)\ge c_2 n^{-2/p}N^{2/p-1}.
$$ 
\end{theorem}

\begin{proof} Let $c_0$ be the constant from Lemma \ref{lem:5}, and let 
\begin{equation}
\label{I1}
c_1=c_0/\sqrt{12}.
\end{equation}
By assumption,
\begin{equation}
\label{I2}
 c_0 \sqrt{N}\le n \le c_1 N .
\end{equation}
We set
\begin{equation}
\label{D7}
L=N, \quad  \ell=\lceil 2c_0^{-2}n^2N^{-1}\rceil, \quad \ell'=\ell+1.
\end{equation}
It follows from (\ref{I2})  that $\ell\ge 2$. Moreover,
from (\ref{D7}), 
\begin{equation}
\label{D8}
n\le c_0\sqrt{\ell N/2}
\end{equation}
and, taking into account that $\ell\ge 2$,
$$ 
\ell/2 \le\ell-1<  2c_0^{-2}n^2N^{-1},
$$
hence, by (\ref{I1}) and (\ref{I2}),
\begin{equation}
\label{D10}
\ell+1\le 3\ell/2< 6c_0^{-2}n^2N^{-1}\le 6c_0^{-2}c_1^2N= N/2.
\end{equation}
We have, by  (\ref{D8}), (\ref{D10}) and (\ref{D7}),
\begin{equation}
\label{D11}
n\le c_0\sqrt{\ell N/2} \le c_0 \min_{j=\ell,\ell+1}\sqrt{j(N-j)}\le c_0\rho(L,\ell,\ell').
\end{equation}
Now we define 
$\psi_j \in L_p^N \quad (j = 0, \dots , L-1)$ 
as
\[
\psi_j (i) = 
  \left\{
   \begin{array}{lll}
(\ell+1)^{-1/p}N^{1/p} & {\rm if} \quad i=j\\
0 & {\rm otherwise.}
   \end{array}
   \right.
\]
We have
$$
S_N\psi_j =(\ell+1)^{-1/p} N^{1/p-1}.
$$
For each $u = (u_0, \dots , u_{L-1}) \in \{0,1\}^L$ define
\begin{equation}
\label{E6}
f_u = \sum_{j=0}^{L-1} u_j \psi_j.
\end{equation}
Since the functions $\psi_j$ have disjoint supports, the system
$(f_u)_{u\in\{0,1\}^L}$ satisfies condition (I). Moreover,
 $f_u\in \mathcal{B}_p^N$ whenever $|u|=\ell,\ell+1$.
Lemma \ref{lem:5}, relation (\ref{D11}) and the left and middle part of (\ref{D10}) 
give
\begin{eqnarray*}
e_n^q(S_N,\mathcal{B}_p^N)&\ge& \frac{1}{2}\min\big\{ |S_N f_u-S_N f_{u'}|\,\big |\, 
|u|=\ell,\, |u'|=\ell+1\big\}\\
&=& \frac{1}{2} (\ell+1)^{-1/p}N^{1/p-1} 
\ge\frac{1}{2}(6c_0^{-2}n^2N^{-1})^{-1/p}N^{1/p-1}\\ 
&=& \frac{c_0^{2/p}}{2\cdot 6^{1/p}}n^{-2/p}N^{2/p-1}.
\end{eqnarray*}
\end{proof}
\section{Comments}
Let us first mention that there remains another gap in the order of
the quantity $e_n^q(S_N,\mathcal{B}_p^N)$ in all the results of
Theorems \ref{theo:3}, \ref{theo:4}, and Corollaries \ref{cor:2}, \ref{cor:3}, namely, the 
region $c_1 N\le n< N$. As we mentioned before, we have $e_n^q(S_N,\mathcal{B}_p^N)=0$
for $n\ge N$ (classical computation of the sum). Hence filling this gap means 
determining how fast 
$e_n^q(S_N,\mathcal{B}_p^N)$ goes to zero in the region close to classical computation.
We did not consider this problem further. It is theoretically interesting, but one should
also mention that its solution would
not say much about the speed-up due to quantum computation: With an effort, just
by a constant factor higher, the problem can be solved with the same error
(in fact, even up to any needed precision) by classical computation.

Finally, we discuss the cost of our algorithm in the bit model of computation.
Here we assume that both $N$ and $n$ are powers of two. The algorithm behind Proposition
\ref{pro:1} and Corollary \ref{cor:1} needs  $\mathcal{O}(nm'')$ quantum gates 
(see Nielsen and Chuang, 2000, Chapter 4, for basics on quantum gates),         
$\mathcal{O}(m'')$ qubits, and makes 
$\mathcal{O}(n^2N^{-1}/\max(\log(n/\sqrt{N}),1))$ measurements to reach error 
$\mathcal{O}(2^{\log N-m''})$. The bit cost of the classical computations
is negligible as compared to the number of quantum gates: We need  
$\mathcal{O}(n^2N^{-1}m'')$ classical bit operations to sort out the wrong
elements and to add the right ones.
The bit cost of the algorithm in connection
with Proposition \ref{pro:2} was already analyzed in Heinrich (2001). It amounts
to $\mathcal{O}(n\log N)$ quantum gates, $\mathcal{O}(\log N)$ qubits,  and 
$\mathcal{O}(k\log k)$ (which is
$\mathcal{O}(\log n\log \log n)$) measurements. The number of classical bit 
operations is $\mathcal{O}(\log n\log\log n\log N)$, and thus, again 
dominated by the number of quantum gates.
Summarizing this for the algorithm
of Theorem \ref{theo:1}, we see that we can implement it with 
$\mathcal{O}(n\log N)$ quantum gates, on $\mathcal{O}(\log N)$ qubits,
 and with
$$
\mathcal{O}(n^2N^{-1}/\max(\log(n/\sqrt{N}),1)+ \log (N/n)\log \log (N/n))
$$ 
measurements. Thus the quantum bit cost differs by at most a logarithmic
factor from the quantum query complexity.


\begin{thebibliography}{BHMT00}

\bibitem{AW:99}
D.~S. Abrams and C.~P. Williams (1999):
\newblock Fast quantum algorithms for numerical integrals and stochastic
  processes.
\newblock Technical report, http://arXiv.org/abs/quant-ph/9908083.

\bibitem{BBC:98}
R.~Beals,  H.~Buhrman, R.~Cleve, and  M.~Mosca (1998):
\newblock Quantum lower bounds by polynomials, 
\newblock Proceedings of 39th IEEE FOCS, 352-361, see also
 http://arXiv.org/abs/quant-ph/9802049.

\bibitem{BHM:00}
G.~Brassard, P.~H{\o}yer, M.~Mosca, and A.~Tapp (2000):
\newblock Quantum amplitude amplification and estimation.
\newblock Technical report, http://arXiv.org/abs/quant-ph/0005055.

\bibitem{EHI}
A. Ekert, P. Hayden, and H. Inamori (2000): Basic concepts in quantum computation.
See http://arXiv.org/abs/quant-ph/0011013.

\bibitem{G1} 
L. Grover (1996):
A fast quantum mechanical algorithm for database search. 
Proc. 28 Annual ACM Symp. on the Theory of Computing, 212--219, ACM Press New York.  
See also http://arXiv.org/abs/quant-ph/9605043. 
 
\bibitem{G2} 
L. Grover (1998):
A framework for fast quantum mechanical algorithms.
Proc. 30 Annual ACM Symp. on the Theory of Computing, 53--62, ACM Press New York. 
See also http://arXiv.org/abs/quant-ph/9711043. 

\bibitem{Gr}
J. Gruska (1999):
Quantum Computing.
McGraw-Hill, London.

\bibitem{He1}
S.\ Heinrich (2001): 
Quantum summation with an application to integration. 
Submitted to J. Complexity. 
See also http://arXiv.org/abs/quant-ph/0105116. 

\bibitem{He2} 
S.\ Heinrich (2001a): 
Quantum integration in 
Sobolev classes (in preparation).

\bibitem{HN1} 
S. Heinrich and E. Novak (2001): Optimal summation and integration by deterministic, 
randomized, and quantum algorithms, submitted to the Proceedings of the 4th International
Conference on Monte Carlo and Quasi-Monte Carlo Methods, Hong Kong 2000 (to appear).
See also http://arXiv.org/abs/quant-ph/0105114

\bibitem{NW} 
A. Nayak and F. Wu (1999):
The quantum query complexity of approximating the median and related statistics. 
STOC, May 1999, 384--393, 
see also http://arXiv.org/abs/quant-ph/9804066.

\bibitem{Nie}
M. A. Nielsen and I. L. Chuang (2000):
Quantum Computation and Quantum Information, Cambridge
University Press.

\bibitem{N2} 
E. Novak (1988):  
Deterministic and Stochastic Error Bounds in Numerical Analysis. 
Lecture Notes in Mathematics {\bf 1349}, Springer. 

\bibitem{Nov95} 
E. Novak (1995):
The real number model in numerical analysis. 
J. Complexity {\bf 11}, 57--73. 

\bibitem{Nov01} 
E. Novak (2001):
Quantum complexity of integration.
J. Complexity {\bf 17}, 2--16.
See also http://arXiv.org/abs/quant-ph/0008124. 

\bibitem{P} 
 A. O. Pittenger (1999):
Introduction to Quantum Computing Algorithms.
Birk\-h\"auser, Boston.

\bibitem{S3}
P. W. Shor  (2000):
Introduction to Quantum Algorithms. 

See http://arXiv.org/abs/quant-ph/quant-ph/0005003.
  
\bibitem{TW} 
J. F. Traub and H. Wo\'zniakowski  (2001):
Path integration on a quantum computer (in preparation).

\bibitem{TWW} 
 J. F. Traub, G. W. Wasilkowski, and H. Wo\'zniakowski  (1988):
Information-Based Complexity. Academic Press. 
\end{thebibliography}
\end{document}